\documentclass[%
 reprint,
 amsmath,amssymb,
 aps,
floatfix,
]{revtex4-2}

\usepackage[english]{babel}
\usepackage{graphicx}% Include figure files
\usepackage{dcolumn}% Align table columns on decimal point
\usepackage{bm}% bold math
\graphicspath{ {./figures/} }
\usepackage{cleveref}
\usepackage{chemformula}
\usepackage{array,multirow,graphicx}

\usepackage{amssymb}
\usepackage{amsbsy}
\usepackage{amsmath}
\DeclareUnicodeCharacter{2212}{-}
\begin{document}

\preprint{APS/123-QED}

\title{From mixing to displacement of miscible phases in porous media: The role of heterogeneity and inlet pressures}% Force line breaks with \\

\author{Yahel Eliyahu-Yakir}
\author{Ludmila Abezgauz}
\author{Yaniv Edery}%
\email{yahel.eliyahu-yakir@epfl.ch }
\email{yanivedery@technion.ac.il}
\affiliation{%
 Technion - Israel Institute of Technology
\\
 EPFL - École polytechnique fédérale de Lausanne,Civil Engineering Institute, LCH
}%

\date{\today}% It is always \today, today,

\begin{abstract}
Miscible multiphase flow in porous media is a key phenomenon in various industrial and natural processes, such as hydrogen storage and geological carbon sequestration. However, the parameters controlling the patterns of displacement and mixing in these flows are not completely resolved. This study delves into the effects of heterogeneity and inlet pressure on mixing and displacement patterns of low-viscosity miscible phase invasion into a high-viscosity resident phase, that is saturating a porous medium. The findings highlight the substantial influence of inlet pressures and heterogeneity levels in transitioning from uniform to fingering patterns at the pore scale. These phenomena are detectable at the Darcy scale, and their transition from a uniform front to finger formation is effectively quantified through a modified Sherwood number. This quantification links microscale patterns to physical properties like velocity distribution, diffusion, and viscosity contrasts. Additionally, the study employs breakthrough curve (BTC) analysis to illustrate the role of higher heterogeneity and inlet pressure in broadening the fluid velocity distribution, leading to the fingering pattern. These research insights provide a non-dimensional approach for future models of miscible phase flow in porous media, linking pore-scale dynamics with macro-scale Darcy-scale observations.
% \begin{description}
% Secondary publications and information retrieval purposes.
% You may use the \texttt{description} environment to structure your abstract;
% use the optional argument of the \verb+\item+ command to give the category of each item. 
% \end{description}
\end{abstract}

%Use showkeys class option if keyword
\maketitle

%\tableofcontents

\section{\label{sec:level1}introduction\protect\\ }

Miscible multiphase flow in porous media is a process where a resident phase is displaced by a different phase within the confinement of a porous structure, and while the phases are different in their characteristics (e.g., density, viscosity, etc.), they can still mix and form an intermediate phase that has characteristics different from both phases. This process is omnipresent in many processes, such as water remediation, hydrogen storage in the subsurface, moisture and solute transport in soils and geological carbon sequestration (GCS)  \cite{seawater_aquifer, heinemann2021enabling, krevor2023subsurface, Glass1989crop, pan2021underground}. In these processes and others, a less viscous phase displaces a resident phase, resulting in increased mobility and causing a fingering pattern due to phase interface instability known as "viscous fingering" \cite{tan1986stability}. Specifically for hydrogen storage and CGS in brine aquifers, the mixing then leads to the acidification of the brine and potential alterations in rock formations and porosity \cite{pan2021underground, moosavi2019influence, leakageVSdepth, Deng_Acid, edery2021feedback, ellis_2011_Acid,Darcy_simulation_oil, Darcy_simulation_water}.  

The main bulk of miscible multiphase flow research was done on 2D Hele-Shaw cells (HS), where a resident phase residing between two circular plates is displaced by the less viscous miscible phase introduced in the plate center \cite{homsy1987viscous, Paterson,king1987fractal,chen1987radial}. As the invasion proceeds, the interface between the phases is ever-growing, due to the circular nature of the invasion, thus reducing the invading phase fluid velocity in the interface. This invasion pattern leads to the said “viscous fingering,” where instabilities in the form of small inhomogeneity on the plate lead to a variation of velocity on the phases interface, which is acerbated by the phases viscosity difference \cite{hopp2019viscous,luo2018particle,pritchard2009linear,pramanik2015viscous}. Furthermore, research has demonstrated that adjusting the gap of the HS cell can regulate instability, and not every gap has the potential to enhance VF dynamics. Consequently, the cell gap significantly influences VF dynamics \cite{HS_gap2022}. However, studies on miscible phase flow in porous media indicate that the formation of fingering patterns is also influenced by the flow rate \cite{bacri1991three,paterson1981radial, slobod}, and the permeability heterogeneity \cite{nicolaides2015impact}. Therefore, while HS cells highlight the role of viscosity difference among the phases, which is the dominant feature at the volume scale, the porous material highlight the pore structure on the micron scale. Studies on immiscible phase flow at the pore scale showed the dominance of the heterogeneous structure and how it impacts the uneven advancement of the invading phase due to capillary or viscous forces \cite{Rabbani, Zhao, levacheANDbartolo, berkowitz_anomalous, Darcy_simulation_water}. While immiscible phases do not mix and lack the resulting viscosity range associated with miscible phase mixture, they do underscore the relevance of pore scale heterogeneity on the  fingering pattern during invasion. Moreover, studies performed on miscible phases that do mix showed that the fingers establish the interface between the phases where mixing occurs \cite{gopalakrishnan2017relative, Birendra}. As such, rigorous experimental investigation exploring the effect of porous media heterogeneity, fluid flow rate, and miscible phases viscosity, on fingering pattern and mixing at the pore scale is currently missing and is crucial for accurate modeling efforts \cite{kim2023miscible, yuan2022new,lei2021pore} 
This research aims to investigate how the heterogeneity of porous media influences mixing and invasion patterns under different inlet pressures and heterogeneity levels, and bridge the gap between the pore scale and the Darcy scale for miscible phase flow in porous media. Using a 2D porous media model, we examine how low-viscosity fluids interact with high-viscosity fluids at the pore scale, under varying heterogeneity and inlet pressures while monitoring the flow rates. Our findings reveal that both inlet pressures and heterogeneity levels significantly impact fingering patterns, resulting in distinct displacement and mixing behaviors. These pore-scale phenomena are reflected at the Darcy scale through flux measurements, demonstrating that they have broader implications for miscible phase flow in porous media. 

\section{Methods and materials}
\begin{figure}[ht] % 4 flow cells
\centering
\includegraphics[width=\linewidth]{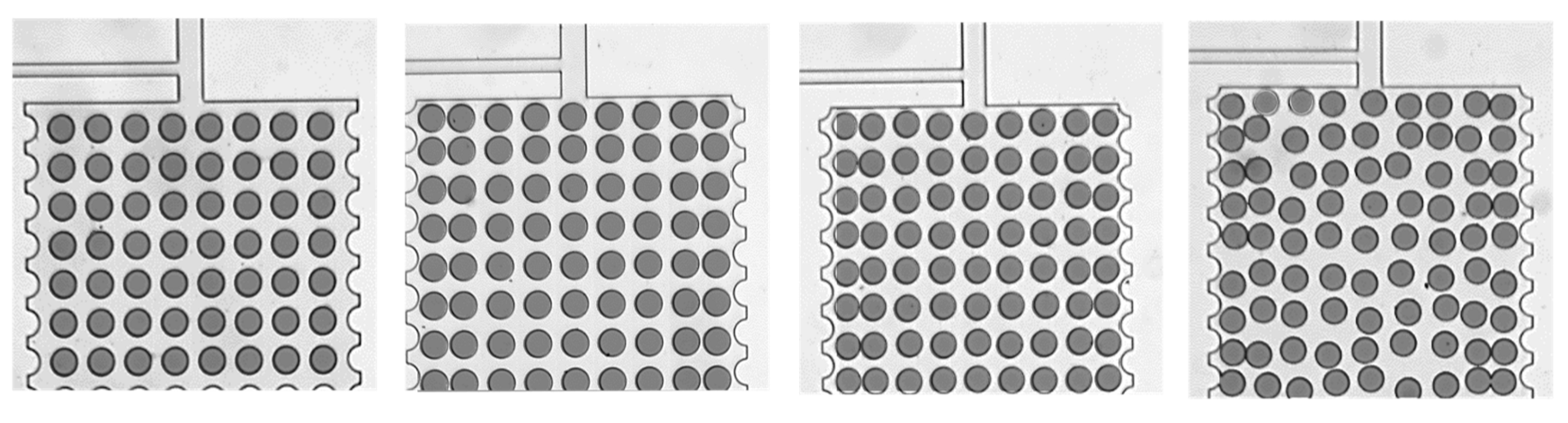}
\caption[flow cells layout]{The 2D flow cells' layout with an obstacle diameter of 50-micrometers. These obstacles are placed on grid points $(x,y)$ of a square lattice, allowing for some disorder with a normalized standard deviation ($\sigma/R$) of 0, 0.01, 0.1, and 0.5 from left to right.}
\label{fig:flowcells} 
\end{figure}

To distinguish between the displacement and mixing within the porous media, we fabricated a 2D porous media with dimensions of 4.5 millimeters in length and 1.3 millimeters in width, and 0.05 millimeters in depth, made out of  Polydimethylsiloxane (PDMS), using a microfluidic mold. We patterned the microfluidic mold with four different heterogeneity levels, where the homogenous pattern is a uniform grid of pillars of 50-micrometer radii, forming an average pore size ($R$) of 50-micrometer. These pillars are placed on grid points (x, y) of a square lattice, allowing for some disorder with normalized standard deviation  ($\sigma/R$) of 0.01, 0.1, and 0.5, forming the heterogeneity levels depicted in \cref{fig:flowcells}. 

We performed a fluid-fluid displacement experiment when the invading fluid is double distilled water (DDW) mixed with Rhodamine 6G, while the defending resident fluid is glycerol and both are inert to the material of the cell. The experimental setup is designed to form a sharp interface between the DDW and the glycerol, with no prior mixing, so as to clearly separate the displacement and the mixing of the fluids. The Rhodamine 6G dye is used as an indicator, which is excited by a 488 [nm] laser and emits at 550 [nm] wavelength. Therefore, as the glycerol is mixed with DDW, the dye is diluted, and the emission amplitude diminishes. The DDW and Glycerol differ in their density ($\rho_{DDW}=1$ and $\rho_{gl}=1.26[gr/cm3]$ respectively) and in their viscosity ($\mu_{DDW}=1\cdot10^{−3}$ and $\mu_{gl}=1[pa\cdot s]$ respectively). For each heterogeneity level, we applied 3 different inlet pressures (15, 30, and 45 [mbar], measured at the inlet), while measuring the inlet discharge for every pressure-heterogeneity configuration. 
In parallel, we developed a MATLAB code to track the invasion of the DDW by the Rhodamine fluorescence and the mixing between phases by the fluorescence decrease due to dye dilution, following a verified linear calibration curve matching the fluorescence amplitude to the phases mixing at each pixel \cite{Rhodamine6g}. Using the  fully saturated and unsaturated images, we directly estimate the thermal white noise error coming from the imaging at less than 1\%. Our experimental configuration entailed monitoring the mixing process using a confocal microscope (Nikon Eclipse Ti2-FP) and a camera (Prime Sigma BSI express) with a frame rate of 0.3 seconds per frame. The flow cell was connected to a pressure pump (Fluigent, with a pressure range of 345 mbar and 69 mbar) to maintain precise and constant pressure while recording the inlet pressure and the discharge changes over time. The reported error by Fluigent for the discharge is 5\%, which is in line with the measurement noise we get.  This monitoring took place through a 1 mm thick borosilicate glass slide (Thermo, Menzel-Glaser),  uniformly coated with PDMS using a spin-coating process, and bounded to the flow cell.

\section{Results and discussion}
The experimental setup, described in the method section, allows us to track the invasion of the DDW through the resident glycerol, and account for the phases of mixing and the invasion pattern.  We will examine the influence of different heterogeneity levels and inlet pressure on three different spatial and temporal aspects: the invading pattern, the mixing pattern, and the overall flux change.
\subsection{Heterogeneity and the invasion pattern}
\begin{figure*}[ht] % raw matrix
\centering
\includegraphics[width=1\linewidth]{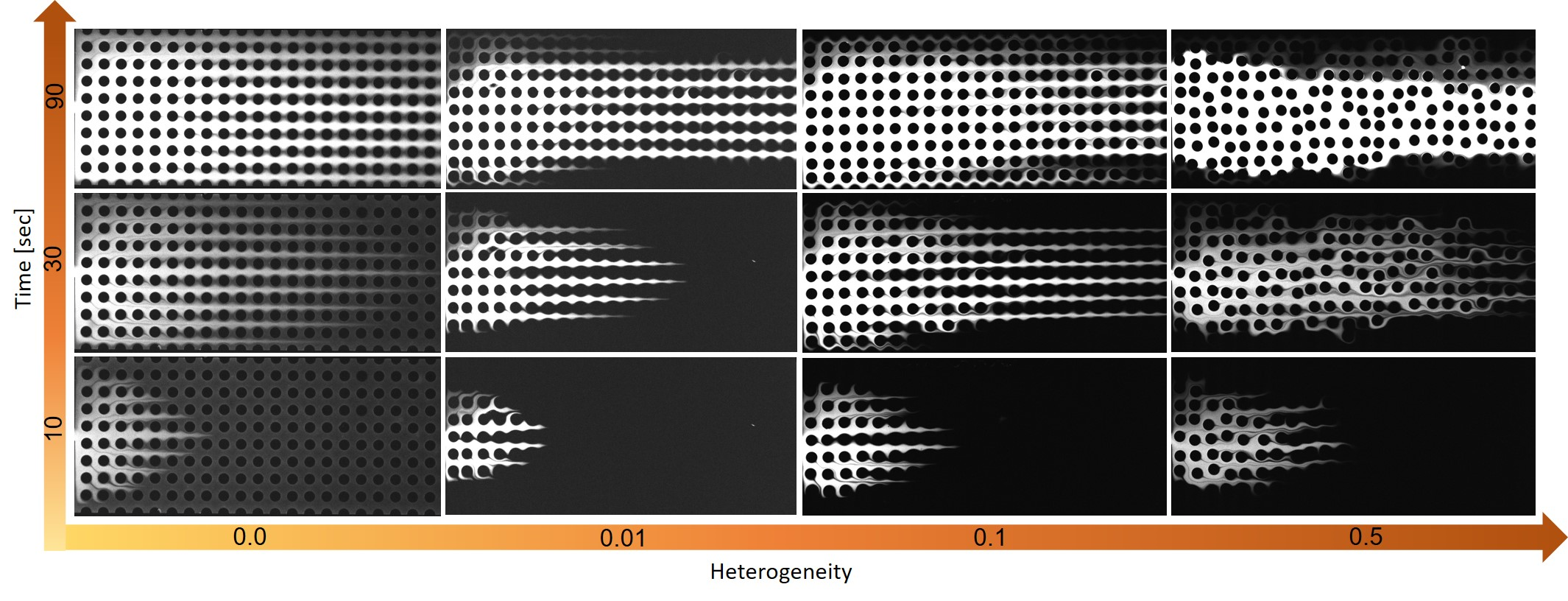}
\caption[Raw data matrix]{Raw data of displacement and mixing pattern variation under a given inlet pressure of 30 [mbar] at 10,30 and 90 [sec] at heterogeneity levels of 0.0, 0.01, 0.1, and 0.5}
\label{fig:raw_matrix} 
\end{figure*}

Pore scale heterogeneity of porous media is the outcome of pore size variation, that dictate the ratio of surface frictional forces on the flowing fluid, while the fluid viscosity, determine the magnitude of these frictional forces. Through these frictional forces, the heterogeneity in the flow cell causes variation in the local velocity due to the connectivity among pores and their sizes along the cell thus forming a preferable path with less overall resistance. The velocity profile, within each pore, is affected by the shear stress that originates from the pore boundaries due to the non-slip condition. Meaning, that the velocity decreases with the pore cross-section, leading to higher resistance. This pore-size resistance mechanism scales with the fluid’s viscosity due to its relationship to the shear stress. Therefore, in a porous media experimental setup, a change in the resistivity is expected compared to the HS cells, yet the coupling between the heterogeneity, resistivity change, and invasion pattern is currently missing. 
We observe that in the homogenous porous medium, the point source inlet for the DDW dictates an initial invasion plume that, over time, evolves into a uniform invasion front despite the point source origin (\cref{fig:raw_matrix}, $\sigma/R=0.0$, with a pressure of 30 [mbar]). This uniform front is different from the mechanism of viscous fingering in HS cell where one would expect a single finger to emerge \cite{slobod}. Keeping the same fluids and pressure head while increasing the heterogeneity level as stated in the method section leads to a change in the invasion pattern of the DDW in the resident Glycerol. The uniform front in the homogenous configuration switches to a single finger that becomes thinner and more pronounced as the heterogeneity increases, captured by the invasion pattern morphology differences between $\sigma/R=0.01$  to $0.1$, and especially $0.5$ In \cref{fig:raw_matrix}. These morphological changes in the invasion pattern suggest that the initial fingering, due to the large viscosity difference, is secondary to the porous media heterogeneity.
Specifically, in our experiment, the less viscous fluid reduces the resistance and increases the velocity at each pore it invades, whether by replacing the resident phase or by mixing with it. Therefore, the fingering pattern is the result of both (1) pore size resistance mechanism and (2) viscosity evolution at the pore scale. These two processes are coupled, initially, the fingering pattern follows the inner structure and then develops due to the viscous differences among the pores. In contrast, the symmetrical inner structure of the homogenous porous media leads to an even pore-size resistance mechanism, and although there are viscosity differences, due to the fluid distribution within the pore structure, the uniform pore pressure suppresses the fingering pattern. Hence, the spatially uniform pore size resistance mechanism damps the viscous fingering pattern for the homogeneous case,  even though the less viscous fluid invades through a point source at the cell center which should encourage the emergence of a single finger. As such, in the homogeneous case, the invading phase forms a uniform front due to the uniform pore sizes, which homogenizes the applied frictional forces on the fluid, and, as a result, the pore pressure over the cross-section. However, as the heterogeneity increases, the pore size variations lead to non-uniform frictional forces and non-uniform pore pressure, leading to velocity variation. This non-uniform invasion pattern leads to variations in viscosity, which further enhances the non-uniform invasion pattern, thus, the heterogeneity rate determines the invading pattern, while the viscosity change enhances the invasion pattern. This further exemplifies the strong impact of the inner structure on the flow pattern.

\subsection{Heterogeneity and the interplay between mixing and displacement}
\begin{figure*}[ht] 
\centering
\includegraphics[width=1\linewidth]{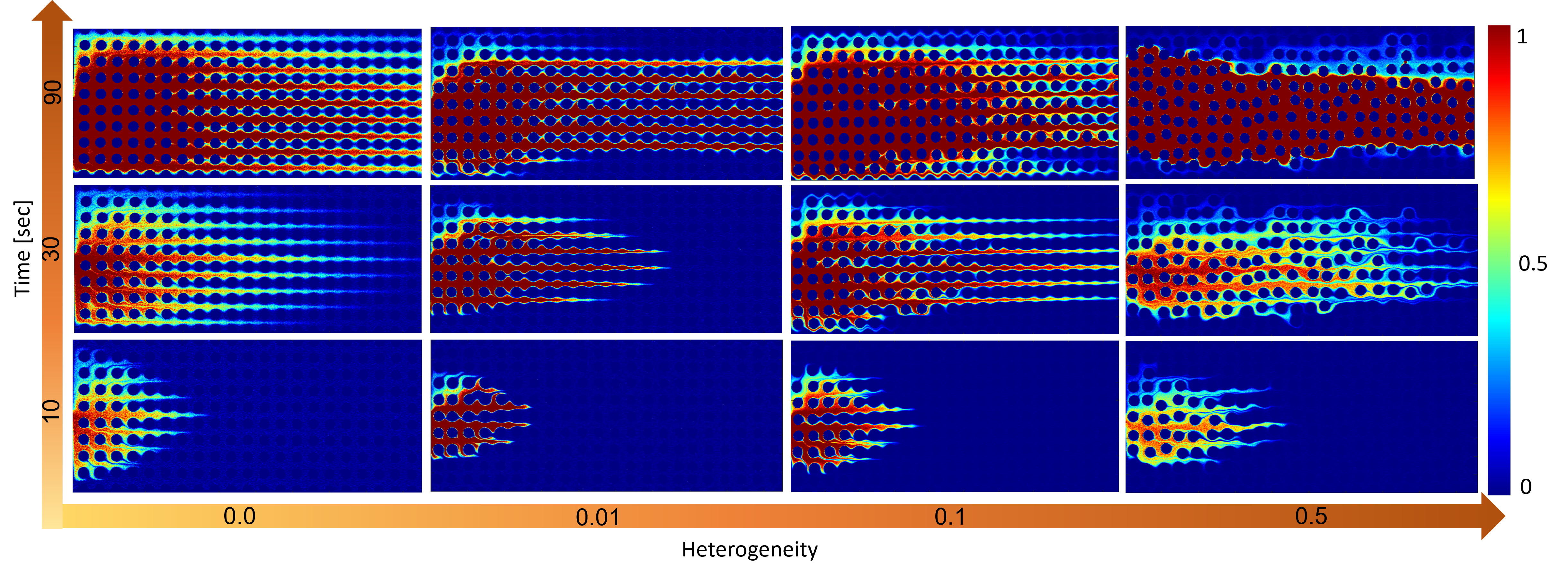}
\caption[Mixing matrix]{An image analysis depicting the local mixing and displacement for flow cells with heterogeneity levels of 0.0, 0.01, and 0.5 for an applied inlet pressure of 30 [mbar] at three different times: 10, 30, and 90 sec. }
\label{fig:hetVStime} 
\end{figure*}

As stated earlier, the observed invasion pattern varies from uniform at the homogeneous case to finger formation as the heterogeneity level increases, yet this observation is focused on the displacement of the fluids, while the effect of pore structure on the fluids mixing is missing. Using a calibration curve on \cref{fig:raw_matrix} the mixing and displacement pattern among the fluids, due to heterogeneity levels, can be observed in \cref{fig:hetVStime}. In the homogeneous cell ($\sigma/R=0.0$), the uniform fluid invasion forms a uniform mixing front with the resident fluid, followed by the displacing fluid. The mixing front maintains the same initial width, with a maximal mixing level in the middle of the front, a remnant to the point source boundary condition. However, as the heterogeneity rate increases ($\sigma/R=0.01$, $0.1$, and $0.5$), the mixing front is less uniform, and a “finger-like" pattern emerge. The displacing front still follows the mixing front, yet at a smaller cross-section within the flow cell; thus, Glycerol remains trapped from both sides of the formed finger. Focusing on the maximum heterogeneity level of $\sigma/R=0.5$, there is a connected finger from the inlet to the outlet within $90$ [sec], while mixing occurs towards the edges of the cell as the finger thickens. As such, the heterogeneity level affects not only the invading pattern but also the mixing pattern, as the main bulk of mixing occurs after the fast finger formation, followed by the finger thickening, in contrast to the homogenous case, where the bulk of mixing occurs in the front of the invasion. 
This finger-like invasion pattern must lead to a complex resistance pattern in time, as it initially follows the pore structure and forms a preferable path with the lowest resistance. Over time, this path remains preferable not only due to the pore structure but also due to the reduced viscosity, and the mixing is reduced to the edges of the finger. Yet this mixing with the trapped Glycerol on the sides must increase the overall fluid viscosity.

\subsection{Discharge change in time as an indicator of spatial displacement and mixing pattern}
\begin{figure*}[ht] 
\centering
\includegraphics[width=1\linewidth]{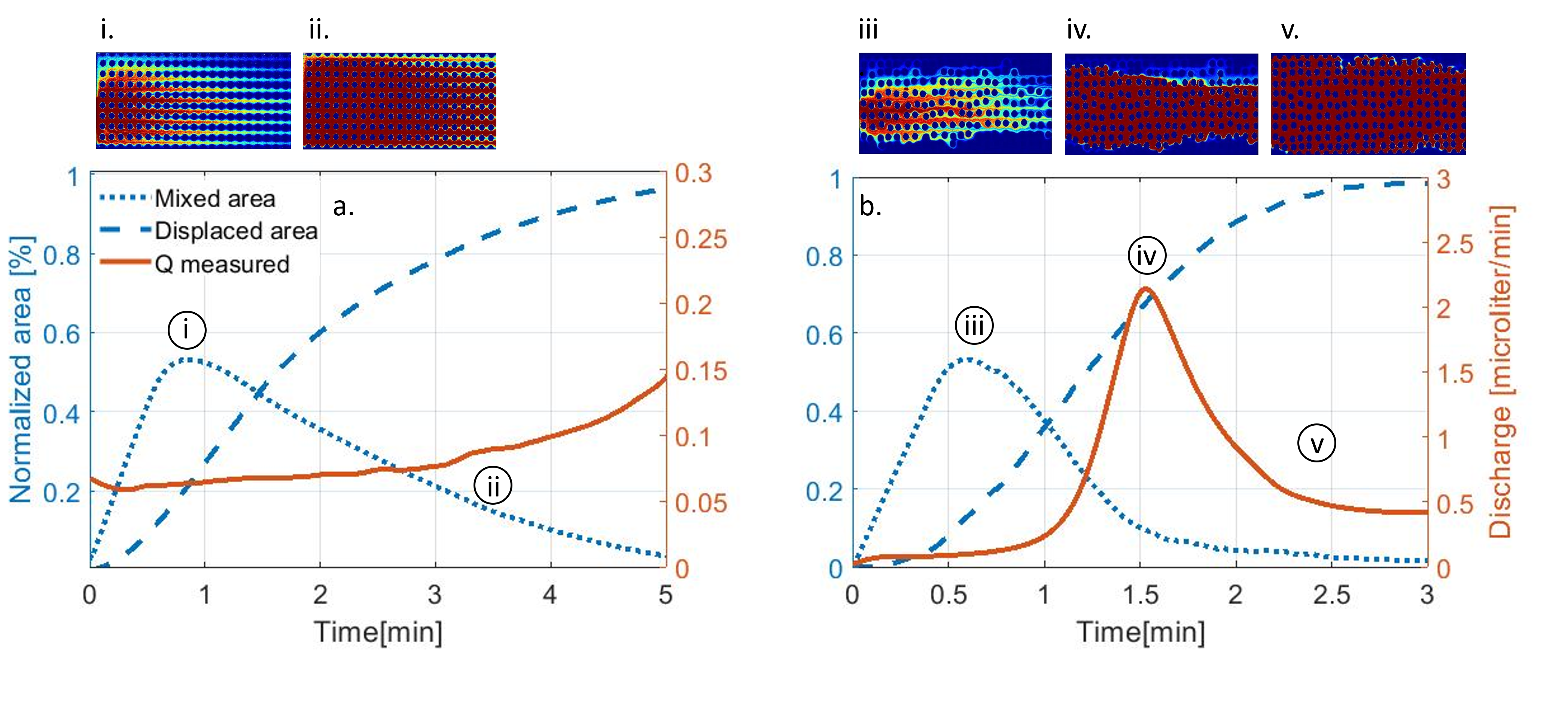}
\caption[Heterogeneous vs. homogeneous in 30 mbar]{a. Homogeneous flow cell b. Heterogeneous flow cell ($\sigma/R=0.0$ and $0.5$, respectively), both with an inlet pressure of 30 [mbar]. The normalized mixing and displacement volumes are represented by the dotted and dashed blue lines, respectively. Discharge with respect to time is marked by the solid orange line, while .a.i. and  a.ii. mark the maximum mixed volume and displaced volume in the cell, respectively,  and b.iii., b.iv., and b.v., mark the maximum mixed volume, the discharge signature for the finger formation, and displaced volume in the cell, respectively}
\label{fig:0.0VS0.5_50mbar} 
\end{figure*}

The pore-scale heterogeneity that causes variation in resistance within the flow cell, leads to changes in viscosity over time and space that must affect the duration required for the invading fluid to reach complete saturation. This change in the displacement time is indicative of the invading phase discharge, as it is basically a change of the volume in time. Specifically, the discharge (Q) must increase in time since the fluid viscosity is reduced by both the displacement and the mixing, thus reducing the cell resistance to flow. However, as the invasion pattern changes with the different heterogeneity and inlet pressures, as indicated in the previous section, so will the discharge increase rate. 
To understand the coupling between heterogeneity and discharge, we first focus on the discharge change over time for the extreme cases of completely homogenous and very heterogeneous porous structures, marked by the orange line in \cref{fig:0.0VS0.5_50mbar} a and b  respectively. These changes in discharge over time are coupled with the normalized mixing and displacement volume, marked by the dotted and dashed blue lines, respectively, in \cref{fig:0.0VS0.5_50mbar} a and b. For the homogeneous case in \cref{fig:0.0VS0.5_50mbar}a, the discharge slope is constant until 200 sec, because of the high glycerol content in the front of the cell that moderates the resistance change. However, DDW mixing in the interface constantly reduces the overall fluid viscosity and therefore, flow cell resistance. Furthermore, the displaced fluid behind the mixed area further decreases the resistance to the flow, which leads to a constant increase in the discharge. When the DDW front reaches the end of the cell, most of the cell is saturated with DDW, (point ii), which causes the resistance of the cell to decrease dramatically, and as a result, to a rapid increase in the discharge. In contrast, for the heterogeneous cell in \cref{fig:0.0VS0.5_50mbar}b, the invading phase flows in a preferable path, or ”finger”. When the “finger” reaches the end of the cell, the discharge is at its maximum, as the resistance within the preferable path is at its lowest due to the viscosity drop, and due to the small cross-section of the finger with a limited surface area of the solid which further reduces the shear force. After the finger formation the discharge drops due to mixing towards the edges of the cell, from point iv to point v, and this mixing phase leads to the saturation of the entire cell. As such, the peak in Q(t) indicates a “finger” pattern in time, which has an initial displacement followed by a mixing phase. Therefore, the external discharge measurement provides a signature to the displacement and mixing mechanisms presented in the previous section. Moreover, it is clear that the same mechanism that leads to the change in local velocity, due to the pore structure heterogeneity and viscosity change in time, also affects the flux measurements, which is the spatial integration of local velocities over the cross-sectional area. However, this measurement differs in the fact that it is made over the whole flow cell on the macro scale and can be considered a Darcy measurement. Therefore, the pore scale mechanism that leads to the change in invasion pattern has a clear signature at the Darcy scale, which can be measured externally, and is clearly linked to the velocity distribution at the pore scale. 

\subsection{Transitioning from front to finger invasion}
As shown in the previous section, the level of heterogeneity plays a significant role in shaping the way fluids displace and mix within a cell, primarily by influencing the velocity field. Alongside heterogeneity, inlet pressure also exerts an impact on the scaling of the velocity field. Specifically, the applied inlet pressure serves as an indicator of the spatial distribution of velocity and can be externally scaled by modifying the inlet pressure value. To further investigate the influence of inlet pressure, we conducted identical experiments across all heterogeneity configurations, using three distinct inlet pressure values: 15, 30, and 45 [mbar]. These results are presented in \cref{fig:phaseDiagram} as a phase diagram, illustrating how each of these variables (pressure and heterogeneity level) contributes within a matrix configuration, facilitating an examination of their respective influences. 
As was shown previously, the displacement within the $\sigma/R = 0.5$ flow cell is faster to reach full saturation compared to $\sigma/R = 0.0$, for an inlet pressure of 30 [mbar]. This relative saturation time is still the case for the inlet pressures of 15 and 45 [mbar], and the displacement of the $\sigma/R = 0.0$ and the nearly homogenous case of $\sigma/R = 0.01$ flow cells takes the longest time to be fully saturated by the DDW, while maintaining a uniform displacement with no fingering, as reflected through the discharge rate (Q(t)) measurement pattern. Moreover, the single-finger formation, with its unique ”peak” pattern in the Q(t) measurements, also exists for $\sigma/R = 0.1$, in addition to $\sigma/R = 0.5$ where the heterogeneity level dominates. As the inlet pressure increases, the displacement rate increases, and the peak in the discharge rate appears earlier for the heterogeneous case, forming the connected path between the inlet and outlet faster. Yet as the inlet pressure increases, the fingering pattern also forms for $\sigma/R = 0.01$, where the pore structure configuration is very nearly homogeneous (recall \cref{fig:flowcells}). Apparently, the increase in inlet pressure broadens the narrow velocity distribution by favoring the local viscosity variations, leading to variations in shear forces. The broadaning of the velocity distribution leads to relatively high velocities which shifts the invasion towards a fingering pattern, even though the permeability along the cross-section is almost uniform. In a way, higher pressure increases the shear forces, and acerbates the small differences in the pore structure, which broadens the velocity distribution; thus, transitioning the nearly uniform flow pattern towards a finger formation, while lower pressures decrease the shear forces and narrows the velocity distribution. For $\sigma/R = 0.0$, there is no evidence for the fingering pattern even at a high inlet pressure, as there are no spatial velocity variations, due to the highly uniform pore structure; nonetheless, this fingering may still occur for the limiting case of $\sigma/R = 0.0$ through a local instability in the invasion, which the local viscosity difference will worsen. 
Fingering patterns have already been observed in simulations of miscible fluid-fluid displacement at homogeneous porous media \cite{lei2021pore}. However, in our experiments at $\sigma/R = 0.0$, we did not observe this finger pattern for this specific cell length. It is important to note that these simulations were performed in 2D geometry, and initial results from a Comsol 2D simulations done in our lab for the homogeneous case, while including the third dimension by the surface area of HS plates as done in \cite{homsy1987viscous},  show that in 2D the displacement pattern is controlled by a single finger. However, performing the same simulation in a full 3D configuration recovered a similar uniform displacement pattern observed in our experiments, marking the importance of experimental observations in this field. We believe that the correction for the added surface area between the 2D and the 3D case is not sensitive enough to miscible phase flow, where the shear forces scale with both pressure and heterogeneity. The mixing at the front of the invading phase, together with the specific viscosity ratio of the invading and residence fluids, restrain the fingering pattern. These results are currently out of the scope of this study and will be presented in a future study comparing the Comsol simulation with the experiment while accounting for the local shear forces. The fingering pattern may also appear if the cell is made longer, which will allow the instability that generates the fingers to overcome the suppressing nature of the uniform shear forces applied in the homogeneous case.
\subsection{The finger formation affect on the breakthrough curve tailing}
As stated in the previous section, the heterogeneity level and the driving pressure lead to the velocity spatial distribution through the shear stress, which in turn leads to the mixing and displacement pattern of the invading miscible phase. However, in our experiments, we cannot directly measure the velocity distribution, yet we do measure the discharge as an indicator for the displacement pattern, which is an accessible Darcy measurement done indirectly in the lab and in the field. Yet there is an additional measurement done in the lab and field scale that allows us to analyze concentration changes of the fluid, originating in the pore scale and propagating to the Darcy scale, namely the breakthrough curve (BTC). Previous studies have related the velocity distribution with the tailing of the BTC \cite{edery2015anomalous,aramideh2018pore,kang2014pore}. These studies indicate that as the velocity distribution in the pore scale increases, so does the tailing increases, and the transport is considered more anomalous \cite{BMB11,bijeljic2013predictions}. Currently, we will only focus on the BTC tailing as an indicator of the broadening of the velocity distribution, while the nature of the anomalous flow will be the subject of a future study. 
In a BTC plot, the normalized concentration of the dye, $C/C_o$ is presented versus the number of pore volumes. This will be done through the measured volume change using the dye concentration image analysis, where $C$ is the volume of the injected fluid, and $C_0$ is the total cell volume. The pore volume (PV) is defined as the cumulative inlet volume divided by the cell's volume, and it is a dimensionless number that scales the time by the time it takes a single volume to be replaced. Scaling the time by the volume exchange time allows us to compare the effect of the dispersion alone as we divide it by the average velocity. As such, the BTC's tailing should be indicative of the velocity field, which is the outcome of the fluid's displacement and mixing patterns.  
\par
\begin{figure*}[ht] 
\centering
\includegraphics[width=1.1\textwidth]{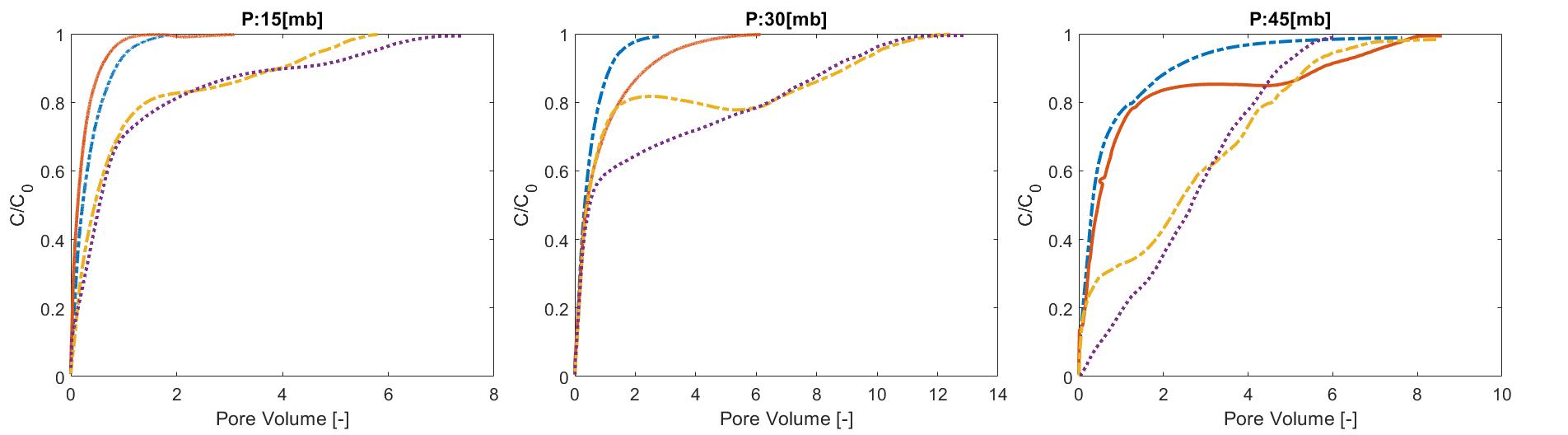}
\caption[Break through curve]{Break through curve for inlet pressure of 15,30 and 45 [mbar] from left to right}
\label{fig:BTC} 
\end{figure*}

In \cref{fig:BTC} 3 different inlet pressures are presented: 15, 30, and 45 [mbar], and with each increase of the pressure, the  differences between the BTC's of various heterogeneity levels becomes more pronounced. At an inlet pressure of 30 [mbar] in \cref{fig:BTC}a, the system can be divided into two groups: (1) the homogeneous group of $\sigma/R=0.0 $ and $ 0.01$ and (2) the heterogeneous group of $\sigma/R=0.1 $ and $ 0.5$, which follow the fingering pattern. It is possible to distinguish the BTC curve within the heterogeneous group (2) into two curve slopes: steep and moderate. The steep slope represents a low PV time needed to establish the finger from the beginning until it is fully developed to the inlet-outlet connection with high velocity. Once the finger is steady, the mixing slowly reduces the resident phase volume as the viscosity increases and lowers the velocity, thus requiring more PV; this is the moderate slope, also known as "heavy tail." The moderate slope represents the high amount of PV that was needed to mix and then replace the remaining glycerol from the edges of the cell.  Figure \cref{fig:BTC} shows that the configuration of $\sigma/R=0.5$ requires the most PV to reach saturation, yet in \cref{fig:phaseDiagram} the configuration of $\sigma/R=0.5$ requires the shortest time to reach saturation, and therefore, we understand that the velocity within the formed finger is the highest. This high velocity leads to a larger velocity polarization, and so, the velocity distribution increases. This broadening of the velocity distribution becomes even more significant as the inlet pressure increases, as shown at 45 [mbar] \cref{fig:BTC}c. At this pressure, each heterogeneity level has its own slope, portraying how the increase in the inlet pressure broadens the velocity distribution that affects the pattern of the invading phase, even if the heterogeneity level is only slightly different from the homogeneous configuration. This observation has already been discussed in the context of discharge rate measurement in \cref{fig:phaseDiagram}, supporting the same findings discussed here. Alternatively, as the inlet pressure decreases to 15 [mbar], the velocity distribution becomes narrower, resulting in lower differences between heterogeneity levels, as shown in \cref{fig:BTC}a.

\subsection{Using the Sherwood number to differentiate the phase diagram }
\begin{figure}[ht] 
\centering
\includegraphics[ width=1.3\textwidth, angle =90]{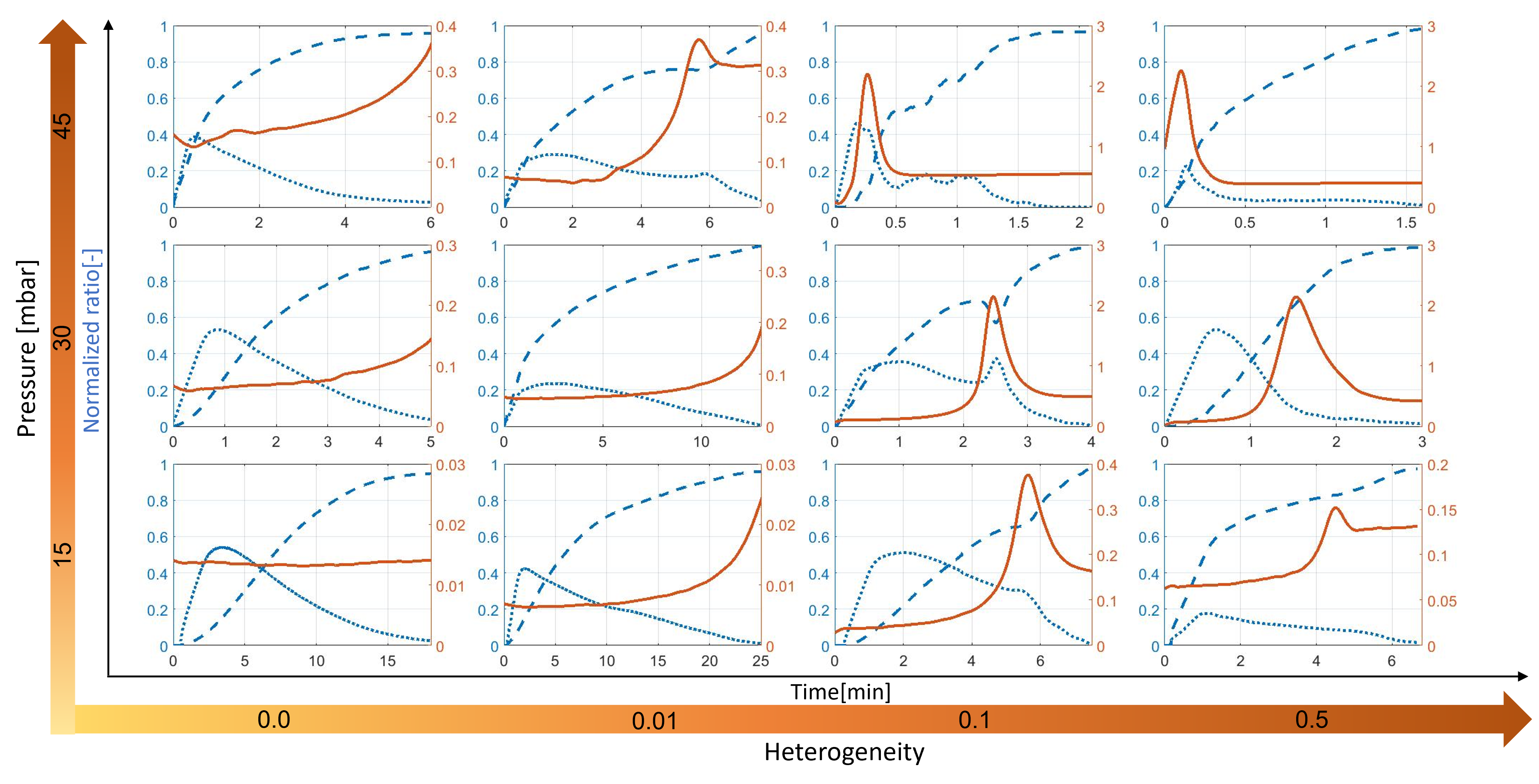}
\caption[Final matrix: All pressures and heterogeneities]{Phase diagram illustrates how heterogeneity and inlet pressure simultaneously influence on the displacement (brackish blue line), the mixing (dotted blue line), and the discharge rate pattern (solid orange line)}
\label{fig:phaseDiagram} 
\end{figure}

Previous sections analyzed the marked transition between the uniform front and the finger formation, set by the change in pressure and heterogeneity configurations, leads to a change in the invasion and mixing pattern (see \cref{fig:phaseDiagram}). While the BTC measurements point to the role of velocity distribution, which is the outcome of shear stresses induced by the pore structure and pressure difference, as the dominant mechanism for this pattern change. However, another outcome of this velocity distribution, leading to the pattern change, is the change in the convective mass transfer rate for the invasion pattern and the diffusion rate for the mixing, which are the hallmarks of the Sherwood number. The Sherwood number is suitable in our experimental context as the mixing among the phases must follow the diffusion of both the water phase through the glycerol and the glycerol through the water over their mutual interfaces; while the water advancement scales the rate of convective mass transfer.  Following this analogy, we define the convective mass transfer ($q$) in our system using the Darcy relation: 
\begin{align}\label{eq:Darcy}
q=-\frac{k}{\mu_{gl}L}\Delta P , 
\end{align}
where $k$ is the permeability, $L$ is the length of the cell, and use the dominant viscosity in our setup, namely $\mu_{gl}$, as it is three orders larger than $\mu_{DDW}$. The permeability is calculated directly by the Kozeny-Carmen equation:
\begin{align}\label{eq:Kozany_carmen}
k=\frac{\phi^3}{cT^2A^2(1-\phi)^2}, 
\end{align}
where $\phi$ is the porosity, $c$ is the spherical coefficient, here equal to 1 as the pillars are round, $T$ is the tortuosity, and $A$ is the hydraulic radii defined as the surface, $S$, to volume, $V$  ($A=\frac{S}{V}$). The tortuosity can be directly calculated from the normalized standard deviation $\sigma/R$, which marks the range for the pillar center movement from a uniform grid using the following relation, $T=1+\sigma/R$.  As the interface between the phases changes due to the torturous path, we define an effective diffusion coefficient $D_{eff}=\frac{D}{T}$, which scales the diffusion with the tortuosity, as shown in many studies (\cite{kim1987diffusion}, \cite{quintard1993diffusion},  \cite{quintard1993transport}, and \cite{beyhaghi2011estimation}). This derivation provides the Sherwood number calculated from the pore structure and marked by the updated permeability and tortuosity:
\begin{align}\label{eq:basic_sher}
Sh=\frac{q} {D_{eff}/R}=\frac{R\phi^3 \Delta P }{TA^2(1-\phi)^2\mu_{gl}L D},
\end{align}
where $R$ is the mean pore size. As stated earlier, the low-viscosity fluid invades through the low-resistance pore, which further reduces the overall resistance by the viscosity change, leading to higher velocity. Therefore, this equation considers the geometry of the flow cell, yet it does not consider the change in the velocity distribution due to the change in viscosity as the invasion occurs. For that, we approximate the velocity distribution variance by the pore structure variance, as the latter leads to the shear forces that scale the velocity distribution. We then add to the Sherwood number (\eqref{eq:basic_sher}) the impact of the high-velocity values, as marked by one normalized standard deviation, and combined with the viscosity contrast ($\mu_{C}= \frac{\mu_{gl}-\mu_{DDW}}{\mu_{gl}+\mu_{DDW}}$, shown in \cite{paterson1981radial}):
\begin{align} \label{eq:mod_sher}
Sh_{eff}=\frac{Rq}{D_{eff}}+\frac{q\sigma/R}{D/w\phi}\mu_{C} =\frac{\phi^3 \Delta P }{T^2A^2(1-\phi)^2\mu_{gl}L D}(RT+ \frac{w\phi\mu_{C}\sigma}{R}), 
\end{align}
where $w$ is the width of the cell, marking the crossectional area for the velocity variations. This calculated modified Sherwood number has a marked transition around  $Sh_{eff}=0.05$, as depicted in table \ref{tab:Sherwood}, and above this value the invasion pattern follows a finger formation, while below this value the invasion pattern follows a uniform front, as seen in \cref{fig:phaseDiagram}. Thus, using only known and measurable quantities in the experimental setup while approximating the  broadening of the velocity distribution by the added normalized standard deviation, the observed pattern transition between the finger to the uniform front is captured through the modified Sherwood number in a phenomenological way.  
\begin{table}
    \centering
    \begin{tabular}{|c|c|c|c|c|c|}
\hline
    \multicolumn{1}{|c|}{} & \multicolumn{5}{c|}{$\sigma/R$}                                             \\ \hline
\parbox[t]{4mm}{\multirow{3}{*}{\rotatebox[origin=c]{90}{P [mBar]}}}   & & 0.0 & 0.01 & 0.1 & 0.5\\ \cline{2-6} 
                           & 45 &  0.033 & 0.064 & 0.21 & 0.53 \\ \cline{2-6} 
                           & 30 & 0.022 & 0.043 & 0.15 & 0.35 \\ \cline{2-6} 
                            & 15 & 0.011 &  0.021& 0.073 & 0.18 \\ \cline{2-6} 
                            \hline
   \end{tabular}
    \caption{The table depicts the modified Sherwood,  by calculating the permeability directly from equation \eqref{eq:Kozany_carmen}, using the known flow cell geometry, and calculating the flux from equation \eqref{eq:Darcy} using the permeability, applied pressure, and a single normalized standard deviation for the flux as shown in equation  \eqref{eq:mod_sher}. The table shows a marked transition around  $Sh_{eff}=0.05$, where above it there is a finger formation, and below it, there is a uniform front, inline with \cref{fig:phaseDiagram}}
    \label{tab:Sherwood}
\end{table}
\section{Summary and Discussion}
This scientific investigation explores the impact of heterogeneity and inlet pressure on mixing patterns in porous media, using a 2D model to study interactions between low- and high-viscosity fluids. The research reveals that both inlet pressures and heterogeneity levels crucially influence fingering patterns, leading to variations in fluid displacement and mixing behaviors at the pore scale. These phenomena are observable at the Darcy scale through flux measurements featuring the transition between uniform and finger formation. This transition can be captured by a modified Sherwood number that phenomenologically links these microscale patterns to physical properties such as velocity distribution, mixing through diffusion, and viscosity contrasts. This transition is also apparent in the BTCs, which also point to the significant role the velocity distribution has in initiating the finger formation. We summarize our findings as follows:
\begin{itemize}
    \item In the homogeneous porous media, an initially non-uniform invasion due to a point source evolves into a uniform front, contrasting with the expected single-finger pattern in viscous fingering; however, increasing heterogeneity shifts this to a more pronounced, thinner single-finger pattern. This indicates that the invasion pattern is primarily dictated by the media's heterogeneity, with variations in viscosity further enhancing this pattern, underscoring the significant influence of the medium's inner structure on flow dynamics.
    \item In the homogeneous media, a uniform fluid invasion results in a consistent mixing front, while increased heterogeneity leads to less uniform, finger-like patterns with the displacing fluid following a narrower path, leaving glycerol trapped on either side. This finger-like invasion, initially guided by pore structure, evolves over time to favor paths of lower resistance and reduced viscosity, with mixing primarily occurring at the edges of the finger and affecting the overall fluid viscosity.
    \item The pore-scale heterogeneity in porous media affects both the viscosity and resistance within the flow cell, influencing the time taken for the invading fluid to achieve complete saturation and altering the rate of discharge. This discharge pattern, which follows the heterogeneity and inlet pressure, offers a macro-scale Darcy measurement, indicative of the underlying pore-scale processes, and linking changes in local velocity and viscosity with observable changes in fluid flow and invasion patterns at the macro scale.
   % \item Both heterogeneity and inlet pressure significantly influence fluid displacement and mixing within a cell, with inlet pressure affecting the velocity field and displacement patterns across different levels of heterogeneity. Higher inlet pressures intensify the velocity distribution and fingering patterns, especially in nearly homogeneous configurations, while homogeneous cells show uniform displacement without fingering even at high pressures, underscoring the critical interplay of pressure, heterogeneity, and local viscosity variations in shaping flow patterns within porous media.
    \item Both heterogeneity and inlet pressure significantly influence the velocity distribution for miscible phase flow in porous media, affecting the displacement and mixing patterns of the invading fluid phase, as indicated by the breakthrough curve (BTC) analysis. The BTC analysis reveals that higher inlet pressures enhance the velocity distribution, leading to distinct displacement patterns across varying heterogeneity levels, with more pronounced differences in velocity and mixing patterns observed at higher heterogeneity and pressure levels.
    \item The marked transition from a uniform front to finger formation in miscible fluid invasion and mixing patterns within porous media is governed by the velocity distribution, which is influenced by pore structure, and pressure differences. This pattern change is analytically captured by a modified Sherwood number, incorporating factors such as the convective mass transfer rate, diffusion rate, pore structure variance, and viscosity contrast, effectively linking these physical properties to the observed transition between uniform and fingered flow patterns in a phenomenological manner.
\end{itemize}
These findings about the displacement and mixing mechanisms of miscible phase flow in porous media are relevant for improving existing models of miscible multiphase flow, particularly in geological and industrial processes. By demonstrating how fingering patterns and mixing behaviors in miscible phases are influenced by the heterogeneous structure and driving pressure, the research provides insights for refining predictive models used in applications like geological carbon sequestration and hydrogen storage. It highlights the importance of considering both viscosity differences and pore scale heterogeneity, which are missing in 2D Hele-Shaw cell models. The study also points to the limitations of 2D simulations in capturing the additional boundary effect on shear stress, which will be the subject of a future study. These insights are particularly relevant for environmental applications, such as the sequestration of supercritical $CO_{2}$ and its interaction with resident brine, as it ties the pore-scale displacement and mixing with the Darcy-scale observations measured for environmental and industrial processes. 

% \begin{table}
%     \centering
%     \begin{tabular}{|c|c|}
% \hline
%     \multicolumn{1}{|c|}{} & \multicolumn{2}{}{Parameters}                                              \hline
%                            & $L$ & $4.5 [mm]$\\ 
%                            & $W$ & $1.3[mm]$\\ 
%                             & $h$ & $0.05[mm]$ \\
%                            & $\mu_{gl}$ & $1 [pa\cdot s]$\\ 
%                            & $\mu_{DDW}$ & $10^{-3} [pa\cdot s]$\\ 
%                             & $\rho_{gl}$  & $1.26[gr/cm3]$\\
%                            & $\rho_{DDW}$ &  $1 $[gr/cm3]$\\ 
%                            & $D$ & $950 \um m^{2}/s$\\
%                             & $A$ & $65.5   69.26   69.2   71.7 [mm]$\\ 
%                             & $\theta$ & $0.67    0.63    0.63    0.61$\\ 
%                             & $T$ & $1    1.1    1.31    1.7$\\ 
%                             & $k$ & $0.72    0.45    0.31    0.15 \times 10^{-3}[mm^{2}]$ \\ 

%                             \hline
%    \end{tabular}
%     \caption{The table presents the parameters used to calculate $Sh_{eff}$}
%     \label{tab:parameters}
% \end{table}
% \clearpage

\clearpage

\begin{acknowledgments}
We thank Prof. Marco Dentz for his insightful comments and discussions. Y.E. and Y.E.Y thank the support of the Israel Science Foundation (ISF grant No. 801/20).
\end{acknowledgments}

\appendix

\bibliography{bibliography}% Produces the bibliography via BibTeX.

\end{document}